# UNSTABLE FLOW AND NON-MONOTONIC CONSTITUTIVE EQUATION OF TRANSIENT NETWORKS

Eric MICHEL, Jacqueline APPELL, François MOLINO,
Jean KIEFFER, Grégoire PORTE [a]

Groupe de Dynamique des Phases Condensées, Case 026,
Université Montpellier II, 34095 Montpellier Cedex 05, France

**Abstract:**

We have measured the nonlinear rheological response of a model transient network over a large range of steady shear rates. The system is built up from an oil in water droplet microemulsion into which a telechelic polymer is incorporated. The phase behaviour is characterized which comprises a liquid-gas phase separation and a percolation threshold. The rheological measurements are performed in the one phase region above the percolation line. Shear thinning is observed for all samples, leading in most cases to an unstable stress response at intermediate shear rates. We built up a very simple mean field model which involves the reduction of the residence time of the stickers in the droplets due to the chain tensions at high shear. The computed constitutive equation is non-monotonic with a range where *the stress is a decreasing function of the rate*, a feature that indeed makes homogeneous flows unstable. The computed the flow curves compare well to the experiments.

[a] to whom correspondence should be addressed

**Introduction:**

Viscoelastic materials have mechanical properties intermediate between those of elastic solids and viscous liquids (see for instance Ferry *et al.* (1980)): they oppose an instantaneous elastic resistance to a sudden shear strain which decays and eventually vanishes. In most cases, the stress relaxation involves one dominant relaxation time so that, irrespective of their mesoscopic structures (polymer melts, wormlike micelles solutions, transient gels, concentrated suspensions…), their rheological properties in the linear regime can be characterized by two quantities only: the instantaneous shear modulus $G_o$ and the terminal relaxation time $\tau_R$. The situation is more complex in the nonlinear regime at high rates. In many cases (polymer melts, wormlike micelles in high salt..), shear thinning is observed (see for instance Bagley *et al.*(1958) or Menezes and Graessley (1982)), sometimes so strong that the homogeneous flow becomes unstable and shear banding occurs: such situations are well documented in the recent litterature on wormlike micelles at high salt (Schmitt *et al.* (1995), Berret *et al.* (1994) and Lerouge *et al.* (1998) and references therein). In some other cases (associating polymers, charged wormlike micelles in absence of salt, concentrated suspensions..) shear thickening is first observed upon increasing the rate, usually followed at still higher rates by shear thinning (Rehage *et al.*(1986), Liu and Pine (1996), Otsubo (1999), Séréro *et al.*(2000)). These non linear effects are indeed of great practical importance in processings and formulations. However, they specifically depend on the details of the structure and on the particular mechanisms of the stress relaxation and to date they are far from being totally understood.

In the present article, we focuss on the flow behaviour at high shear of a model visco-elastic system obtained from an oil in water (o/w) droplet microemulsion into which a hydrophobically end-modified water soluble polymer is incorporated. The spontaneous tendency of the hydrophobic caps to pin onto the oil droplets linked to one another by the water soluble chains determines the formation of the reversible network (figure 1). And viscoelasticity is observed at sufficient concentrations in droplets and chains. In agreement with the intuitive expectation, the stress response to moderate strains is found almost maxwellian: the instantaneous elastic modulus $G_o$ is controlled by the density of connecting threads and the terminal time $\tau_R$ is determined by the average residence time of the hydrophobic stickers in the droplets. Recent experimental studies (Bagger-Jörgensen *et al.* (1997), Filali *et al.* (1999), Pham *et al.* (1999)) have clarified two important aspects of the physics of such transient networks (Senenov *et al.* (1995)). On the one hand, the telechelic chains actually induce an effective, entropy driven attraction between the droplets (Milner and Witten (1992)) which, for some compositions, leads to a phase separation into a dilute fluid solution and a concentrated transient gel. On the other hand, linear rheology measured upon increasing the amount of telechelic chains have revealed the existence of a percolation threshold below which no viscoelasticity is observed. This notion is quite natural since some finite density of links must certainly be exceeded so that an infinite connected cluster forms, capable of transmitting the torque from one wall of the shear cell to the other. Therefore, the phase diagram of transient networks buit up from telechelic chains comprises a *coexistence line* (which bounds the two phase region) and a *percolation line* (figure 2).

To our knowledge, no systematic study of the flow behaviour of complex surfactant-telechelic polymer systems at high shear rate have been reported to date. But data are available for the flow curves of binary systems consisting of telechelic polymers in solvent (see for instance: Annable *et al.*(1993) and Berret *et al* (2001)). (Such simpler systems also form transient networks since hydrophobic caps in absence of surfactant still self assemble in the form of micelles). They all reveal the following qualitative evolution. Upon increasing the shear rate $\dot{\gamma}$ in steady shear, shear thickening is first seen after the usual Newtonian response when the Deborah number ($\dot{\gamma}.\tau_R$) approaches unity. Several explanations were proposed: i) stiffening of the elastic response of the chains at high elongations as in Berret *et al.* (2001) and/or ii) shear induced enhancement of the connection rate as in Marrucci *et al.* (1993). At still higher rate, shear thinning follows which is usually interpreted in terms of the residence time of the stickers being shortened by the tension of the chains. Although these arguments indeed make sense and in spite of many efforts to build up quantitative models, no theory to date is capable to reproduce all features of the flow curves.

We here report on measurements at high shear rate on our droplets linked by flexible chains. The advantage of the apparently more complex surfactant/polymer mixed system is that the number density of the pinning sites (the droplets) and their average functionality (number of stickers per droplet) can be adjusted independently by fixing the concentration of the microemulsion and the concentration of the telechelic polymer. The composition of the sample can so be moved in the above mentionned phase diagram with respect to the coexistence line and the percolation line. All data reported here are obtained at high enough droplet concentration, far out of the two-phase region, but in the vicinity (above) of the percolation line. All flow curves show similar evolutions with shear thickening followed by shear thinning as described in section I. The shear thickening will be discussed in a later article: it originates from the strain hardening of the elastic modulus. Shear thinning on the other hand, begins with an unstable range where the stress shows erratic fluctuations. At very high rates however, a reproducible robust stress response is recovered. We discuss the shear thinning part of the flow curve in section II on the basis of a toy model which includes the most essential aspect first considered by Yamamoto (1956): namely that the residence time of the sticker is reduced by the tension supported by the chain. The model assumes that the material is homogeneously convected by the flow at all scales. In that sense it is essentially mean field. Qualitatively, the flow curves calculated from the model are comparable to the experimental curves: in particular, the flow instability at intermediate rates is related with the non-monotonic character of the constitutive equation

### Section I: experiments.

*Description of the system*

The investigated system is the one previously described in Filali *et al.* (1999). The o/w microemulsion involves a cationic surfactant –namely Cetyl-Pyridinium Chloride (CPCl)- and the cosurfactant is n-octanol. The droplets are swollen with decane and dispersed in 0.2M NaCl brine so to avoid electrostatic complications arising from the ionic character of the CPCl. The cosurfactant to surfactant weight ratio is fixed the same for all samples (0.25) as well as the oil to amphiphile ratio (0.56). The size and shape of the droplets were characterized in details by neutron scattering (Filali *et al.* (1999)): -spheres of 62Å radius- and were found robust to variations of both the droplet concentration and of the amount of added polymer. The telechelic polymer consists of a $10^4$ MW polyoxyethelene hydrosoluble chain at the two ends of which is grafted an alkyl group C18H37 by the mean of an intermediate isocyanate group. The degree of alkylation (better than 98%) of the end groups was determined by NMR using the method described in details in Hartmann *et al.* (1999).

*Phase coexistence and percolation line*

The droplet concentration is noted $\Phi$: the volume fraction of the hydrophobic cores of the droplets (calculated from the mass fraction of decane and surfactants as reported in Filali *et al.* (1999)). The concentration of telechelic polymer is represented by $r$ which is the average number of hydrophobic stickers per droplet (as calculated from the weight fraction of the polymer, the radius of the droplets 62Å and the volume fraction $\Phi$ of the hydrophobic core of the droplets; see Filali *et al.* (1999) for more details). The phase behaviour represented in figure 2 as function of $\Phi$ and $r$ shows up two phase coexistence at low $\Phi$'s and high $r$'s. This phase separation was discussed in details in Bagger-Jörgensen *et al.* (1997) and Filali *et al.* (1999): it is the signature of the polymer induced attraction between droplets when their average distance is larger than the average size of the polymer coil.

The percolation behaviour was characterized from rheological measurements performed using a RFS-II Reometrics strain controlled rheometer in the cone and plate geometry. The typical stress response of one of our viscoelastic samples ($\Phi$=12.6% and $r$=6) to step shear strain of moderate amplitude is shown on figure 3. Note that the stress response is close to Maxwellian: the fit in figure 3 is obtained with a slightly streched exponential:

$$G(t) = \sigma(t)/\gamma = G(0).\exp(-(t/\tau_R)^{0.9}) \qquad (1)$$

where the exponent (0.9) is very close to unity; this expression provides good fits for most samples considered here (except very close to percolation). The instantaneous shear modulus $G(0)$ and the terminal time $\tau_R$ are so accurately characterized as function of $\Phi$ and $r$. As an example, figure 4 a) and b) show the evolution of $G(0)$ and $\tau_R$ at fixed droplet concentration ($\Phi=12.6\%$) as function of the amount of added polymer $r$. In figure 4 c), the low shear viscosity $\eta(\dot\gamma \to 0)$ of the same series of samples is shown. All three quantities vanish below the threshold value $r_p$ (=3.1 at this droplet concentration) and they show power law evolutions above $r_p$ (see the fits in figure 3):

$$G(0) = G_{00}\left(\frac{r-r_p}{r_p}\right)^{1.79} \quad \text{with } G_{00} = 529\, Pa \qquad (2)$$

$$\tau_R = \tau_{00}\left(\frac{r-r_p}{r_p}\right)^{0.74} \quad \text{with } \tau_{00} = 0.84\, s \qquad (3)$$

$$\eta(\dot\gamma \to 0) = \eta_{00}\left(\frac{r-r_p}{r_p}\right)^{2.5} \quad \text{with } \eta_{00} = 429\, Pa.s \qquad (4)$$

These features are all very similar to those reported in Michel *et al.* (2000) for a similar system. They were indeed discussed at length in term of percolation behaviour: as mentioned in the introduction, viscoelasticity implies that an infinite connected cluster of droplets transmits the shear force from one wall to the other of the shear cell. Systematic rheological measurements all over the one phase region in the phase diagram allowed to draw the tentative percolation line in figure 2 (note that the accuracy is poor at low $\Phi$'s close to the critical point because $\tau_R$ decreases down below the time resolution of the rheometer in that range).

*Non-linear regime: flow curves.*

All data reported below are obtained in the Couette geometry with the same rheometer as above. The viscosity in steady shear as function of the reduced rate (Deborah number: $\dot\gamma.\tau_R$) for the $\Phi=12.6\%$ and $r=3.4$ sample, just above the percolation line, is shown in figure 5 in a semilog plot. This pattern is very typical of transient networks with shear thickening around $\dot\gamma.\tau_R=1$ followed by shear thinning at higher rates (see for instance Annable *et al.* (1993) and Berret *et al.* (2001)). The shear thickening will be studied and commented further in a forthcoming publication where we report on the stress response to step strains of very large amplitude.

Here we rather focus on the shear thinning at high rates. It is more illustrative for this purpose to plot the stress as function of the rate in lin-lin coordinates as displayed for a series of $\Phi=12.6\%$ samples close to and further above the percolation line in figure 6 a. and b. This representation emphasizes the high shear range where the flow pattern is remarkable. For all samples in figure 6, the stress at high rates shows a linear evolution, approximately of the form:

$$\sigma = \sigma_0 + \eta_{eff} \dot{\gamma} \qquad (5)$$

The slope $\eta_{eff}$ increases only slightly with $r$ (in the range 17 to 22 $mPa.s$) whereas the extrapolation to zero, $\sigma_0$ increases roughly linearly with the excess connectivity density above the threshold. At low connectivity ($3.2<r<3.5$), the low slope evolution at high rates connects smoothly to the low rate regime of high effective viscosity. The pattern is different at higher connectivity ($r=3.6$, 3.7 and 3.8 in figure 6 a) and $r = 6$ in figure 6 b)): in between the high shear and low shear ranges (in between the two arrows in figure 6), the stress shows erratic fluctuations with very poor reproducibility from one experiment to the other. Eye examination of the samples sheared at such rates in a specially designed Couette cell with a transparent plexiglass cup revealed: i) wavy deformations of the sample surface at the upper air-fluid interface and ii) progressive, cavitation-like proliferation of bubbles in bulk. All these spurious effects actually vanish and regular flow is recovered both below and above this unstable range. All attempts to improve reproducibility in this range failed and we believe these fluctuations to be the signature of a intrinsic mechanical flow instability of the material. In contrast, the data measured in both the high and low shear ranges are robust and nicely reproducible.

For the sake of comparison, we also collected the flow curve for a droplet sample with $\Phi=12.6\%$ but decorated with a polymer with 5000 MW POE chain grafted at one end only with the same C18H37 hydrophobic sticker. Indeed such a polymer cannot link separate droplets and no network is formed. But the steric and hydrodynamic interactions between droplets are expected to be similar to those between droplets with the polymer with two stickers but forced to loop with both its ends onto the same droplet. This $\Phi=12.6\%$ $r=3.6$ sample with the single sticker polymer shows a Newtonian behaviour up to $\dot{\gamma}=1700s^{-1}$ with a viscosity of $2.1 mPa.s$. This viscosity is indeed smaller, but not so much than $\eta_{eff}$ measured above in the high shear regime of the two sticker samples of similar compositions. This suggests that at very high shear rate *most chains are not elastically active:* most of them form loops with both stickers onto the same droplets. The network no longer forms making the fast flow easier.

**Section 2: Discussion.**

So the main characteristic feature of the flow curves of the transient networks is the instability observed in between the low and the high rates. On the other hand, most interpretations of shear thinning in transient networks are based on the reasonable expectation that the mean residence time $\tau$ of a sticker in a droplet is shortened when the chain is under tension. More precisely, the escape of a sticker is thermally activated and the residence time is determined by the energy of adhesion $E$. In absence of tension, the Arrhenius relation writes:

$$\tau_0 = \nu^{-1} \cdot \exp\left(\frac{E}{k_B T}\right) \tag{6}$$

where $\nu$ is the attempt frequency. Under tension, the adhesion energy is depleted by the reversible work of the tension $\vec{\mathbf{t}}$ on the length $|\vec{\mathbf{a}}|$ of the sticker. So, we expect (Tanaka and Edwards (1992)):

$$\tau(\vec{\mathbf{t}}) = \tau_0 \cdot \exp\left(-\frac{\vec{\mathbf{t}} \cdot \vec{\mathbf{a}}}{k_B T}\right) \tag{7}$$

This will affect the rheological properties in two respects. On the one hand, the terminal relaxation time $\tau_R$ is determined by the residence time $\tau$; since the viscosity of a quasi maxwellian system is of the order of the product of the elastic modulus by the terminal time, a shorter residence time will result in a lower viscosity. This is the first origin of the shear thinning. On the other hand, the number density of active chains will be also modified. Even at rest, not all the chains are elastically active. Some of them loop onto the same droplet (figure 1). Indeed, loops are not stretched by the convective displacement of the droplets and their mean residence time remains unaffected by the shear; only the active chains connecting distinct droplets are submitted to tension and have their residence time biased by the shear. So their relative proportion under steady shear is depleted in the benefit of that of the dead loops. This effect brings a second thinning contribution at high shear.

For a deeper insight into these effects, let us consider the following toy model. It relies on two basic assumptions. First, after each escape, the elastic energy due to the chain tension totally dissipates before the sticker readsorbs onto another droplet: this condition holds as long as the (Zimm) retraction time of the chain is short compared to the inverse shear rate. Second, the probability $P_a$ that an escaped sticker reforms a link rather than a loop is taken constant, independent of the rate. Dead loops and active links have the respective number densities $n_d$ and $n_a$ under shear and $n_{d_0}$ and $n_{a_0}$ in the absence of shear. We assume that the number density $n_f$ of free chains having a sticker dangling free in the solvent is very small compared to both $n_d$ and $n_a$; so we have: $n_d + n_a \approx n$ the total number density of chains (this condition holds when the adhesion energy is high: $E \gg k_B T$; in the present system we expect $E \approx 18 k_B T$). For simplicity, we consider that all links are submitted to the same tension. Under steady shear, the rates of formation of loops and links are counterbalanced by their respective rates of escape. This balance is explicited in the appendix, and one gets:

$$\frac{n - n_{a_0}}{n_{a_0}} = \frac{n - n_a}{n_a} \exp\left(-\frac{\vec{t}.\vec{a}}{k_B T}\right) \qquad (8)$$

The tension $\vec{t}$ indeed arises from the stress which is equally supported by all links. So, forgetting the prefactor, we write:

$$|\vec{t}| = \sigma_{net} / n_a l_0 \qquad (9)$$

Where $l_0$ is a length of the order of the average distance between droplets. On the other hand, the links are elongated by the convective flow by an amount of the order of $\dot{\gamma}.\tau$ where, as stated in (9) $\tau$ depends on the tension $\vec{t}$; so we write:

$$|\vec{t}| = g_0 l_0 \tau.\dot{\gamma} = g_0 l_0 \tau_0.\dot{\gamma} \exp\left(-\frac{\vec{t}.\vec{a}}{k_B T}\right) \qquad (10)$$

where $g_0$ is the stretching modulus of each chain. After some trivial algebra, (8), (9) and (10) give:

$$\frac{a\, g_0\, l_0}{k_B T}.(\dot{\gamma}\tau_0) = \frac{x}{x_0} \ln\left(\frac{x}{x_0}\right) = \dot{\Gamma} \qquad (11)$$

$$\frac{a}{n_{a_0} l_0 k_B T}.\sigma_{net} = \frac{(1+x_0)}{(1+x)} \ln\left(\frac{x}{x_0}\right) = \Sigma_{net} \qquad (12)$$

where the loop to link ratio is noted $x$: $x = (n - n_a)/n_a$ ($x_0$ being indeed the value at rest). So we get parametrized equations for the network stress $\sigma_{net}$ and the reduced rate $\dot{\gamma}\tau_0$. Forgetting the prefactors, we have plotted in figure 7 $\Sigma_{net}$ as function of $\dot{\Gamma}$ fixing the loop to link ratio at rest at the value: $x_0 = 1$ (this value is typically what we expect at the present 12.6% concentration for which the average distance between the droplets is comparable to the polymer size). The curve is non monotonic: the stress increases steeply at low rates, reaches a maximum at $\dot{\Gamma} \approx 5$ and decreases monotonicaly but *very slowly* down to zero at high rates.

But of course, $\sigma_{net}$ in (12) only comprises the elastic contribution of the network. It does not include the purely viscous contribution (sheared solvent and drag of the convected droplets) which certainly dominates at high rates at which $\sigma_{net}$ vanishes. So, for the sake of a qualitative comparison with the experimental data displayed in figure 6, figures 8 a) to c) are obtained after adding to $\Sigma_{net}$ a newtonian contribution of the form $\eta_{eff}\dot{\gamma}$. Figures 8 a) to c) are drawn for increasing "elastic" to "viscous" relative proportions. Actually the calculated curves present many similarities with the experiments. Of course, we find high viscosity at low rates and low viscosity at high rates. In addition, for all three curves, the stress at high rates apparently shows the linear evolution underlined in the experimental section. Moreover, comparing the different patterns in figures 8 provide interesting insights regarding the mechanical instability observed at moderate rates. Figure 8 a) is calculated for a low "elastic" relative proportion: this is what we expect close to percolation where the network is weak. Actually, the evolution from the high slope regime at low shear to the low slope profile at high rate is smooth, very much like that of the $r=3.2$ sample in figure 6 which lies just above the percolation line. Figure 8 b), corresponds to a higher elastic contribution: the stress levels off at a flat point in between the two regimes. This feature compares well to the flow curve of the $r=3.5$ sample. Figure 8 c) is obtained at still higher elastic relative proportion and would correspond to stiffer networks such as $r=3.6, 3.7$ and $3.8$ in figure 6 a) and as $r=6$ in figure 6 b). The calculated curve is strongly not monotonic, with an intermediate range where *the stress is a decreasing function of the rate*. This feature indeed does not show up on the experimental curves: it is well known that non monotonic constitutive equations lead to flow instabilities. In the case of semi-dilute solutions of wormlike micelles, it has been shown conclusively that the non-monotonic constitutive equation results into s shear banded patterns (Cates (1990)). The flow is no longer homogeneous: a viscous and a fluid state coexist, one on the viscous stable branch of the flow curve and the other on the fluid one, the two states being organized in parallel bands stacked along the velocity gradient. At a given rate, the relative proportions of the viscous versus fluid bands is determined by the lever rule. Shear banding in wormlike micelles actually determines a robust stress plateau in the flow curves (Schmitt *et al.* (1994), Berret *et al.* (1994), and Lerouge *et al.* (1998)). In transient networks (in between the arrows in the upper curves in figure 6) however, we do not observe a robust flat stress plateau but rather erratic non reproducible fluctuations of the stress around a roughly constant average value. Nevertheless, we believe these fluctuations to be related to the non-monotonic character of the constitutive equation. Then the question remains open of the absence of steady shear banding for transient networks. A possibility is that off plane fluctuations of the interfaces between bands are amplified rather than damped by the shear: freshly formed bands could be remixed by convection, reform again and so on, leading to chaotic fluctuations of the stress response.

So finally, the model generates flow patterns which qualitatively compare well to the experiments. Attempting more quantitative comparisions would probably not make much sense in view of the approximations made. In particular, all links are assumed to support identical tensions and the tensorial character of the stress is neglected: in reality, the tension depends on the relative orientation of the link with respect to the shear directions. We believe however that this neglected aspect is not essential for the purpose of understanding the gross distinctive features of the flow pattern. The concern is more serious regarding the mean field character of the model: it assumes that the macroscopic affine deformation is homogeneously transmitted at all scales within the network. This is certainly not true close to the percolation where the infinite cluster supporting the stress has a fractal structure up to large scales: more densely connected subclusters are linked together by looser weak parts. Under shear, the deformation indeed concentrates on the weak parts leaving the denser regions almost unsheared. As an example of a spurious effect, the mean field scheme would predict a terminal time $\tau_R$ identical to the residence time $\tau_0$, therefore independent of the connectivity density of the network. This is in clear contradiction with the data in figure 3 b) where $\tau_R$ vanishes as a power law at the percolation point. The mean field picture actually breaks: in a non-homogeneous network, disconnecting the looser regions only is sufficient for a complete release of the stress. So the stress relaxation is faster in a looser network. The mean field approximation has probably no dramatic consequence on the non monotonic character of the constitutive equation which would survive to a more realistic approach. On the other hand, deviations from mean field could be at the origin of the absence of steady shear bands at intermediate shear rates.

**Conclusion.**

We have measured the nonlinear rheological response of model transient networks over a large range of steady shear rates. Shear thinning is observed in all cases with an unstable stress response at intermediate rates for densely connected networks. Following previous theoretical analysis, we interpreted the shear-thinning as the result of the reduction of the residence time due to the tension. In spite of its oversimplifications, the model treats in a consistent way the two main consequences: i) the terminal time is reduced and ii) the link to loop ratio is depleted. The calculated flow curves look similar to the experimental shear patterns: an unstable regime is effectively predicted at high enough connectivity density in the form of a non-monotonic constitutive equation. The absence of a steady shear banded flow probably arises from the heterogeneous distribution of the stiffness of the network.

Another interesting issue is the shear thickening effect at moderate rates. Experiments performed with a different system allowing for denser connectivity indicate that this effect disappears at high connectivity density far above the percolation line. We will report on this matter in a forthcoming publication.

**Appendix**

In order to derive expression (8), we introduce $n_{free}$ the (very small) number density of chains with a dangling sticker, and we consider the rates of formation of links $n_{free} \nu_f P_a$ and loops $n_{free} \nu_f (1 - P_a)$ respectively (where $\nu_f$ is the sticking attempt frequency and $P_a$ is the probability that the attempt results into a link). The escape rates are $n_d \nu_0$ for the loops and $n_a \nu_0 . \exp(\vec{t}.\vec{a} / k_B T)$ for the links. In the steady state, the rates of formation and escape of each species are equal:

$$n_d \nu_0 = n_{free} \nu_f (1 - P_a) \qquad (A1)$$

$$n_a \nu_0 . \exp(\vec{t}.\vec{a} / k_B T) = n_{free} \nu_f P_a \qquad (A2)$$

Combining (A1) and (A2) and noticing that $P_a / (1 - P_a) = n_{a_0} / (1 - n_{a_0})$ one gets immediately (8):

$$\frac{n - n_{a_0}}{n_{a_0}} = \frac{n - n_a}{n_a} . \exp\left( -\frac{\vec{t}.\vec{a}}{k_B T} \right) \qquad (8)$$

**References and notes.**

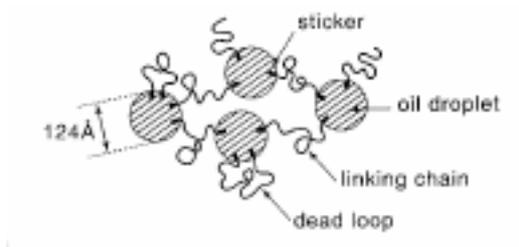

Figure 1 Schematic of the transient network

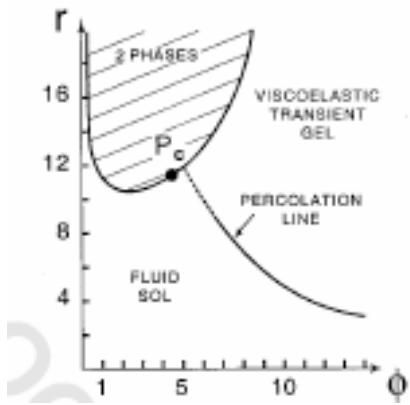

Figure 2 Phase diagram. Note that the accuracy is poor for the percolation line in the dilute range close to the coexistence line (dotted part); it is not clear whether the percolation line ends up at the critical point or slightly off.

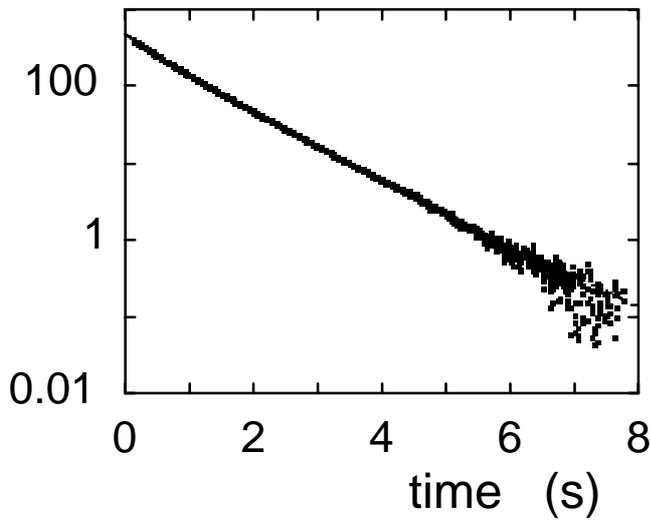

Figure 3): Stress relaxation after a step strain of 20% amplitude for the $\Phi = 13.7\%$, $r = 6$ sample.

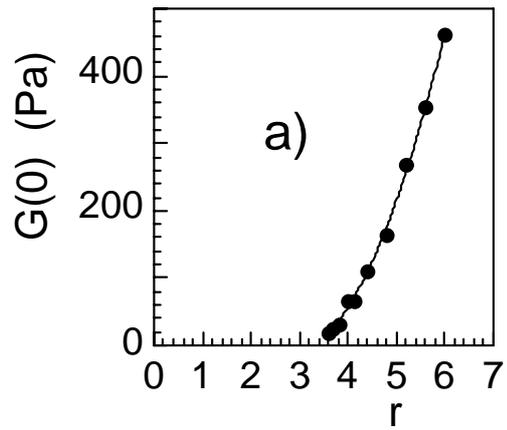

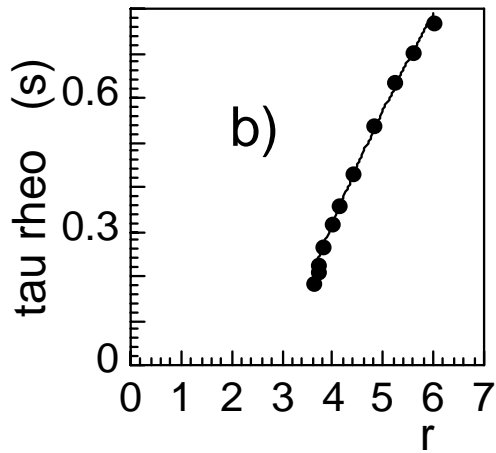

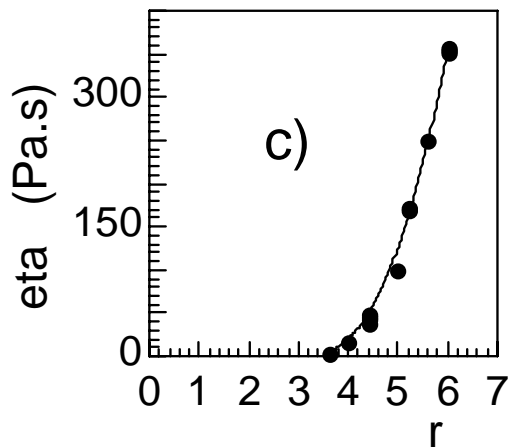

Figure 4: a): $G(0)$, b): $\tau_R$, and c) $\eta(\dot{\gamma}\to 0)$ as function of the connectivity index $r$ for the $\Phi = 13.7\%$ series of samples. The three quantities vanish at the same percolation threshold $r_p = 3.1$: the continuous lines are obtained by fits to the power laws (2), (3) and (4) with the same value for $r_p$.

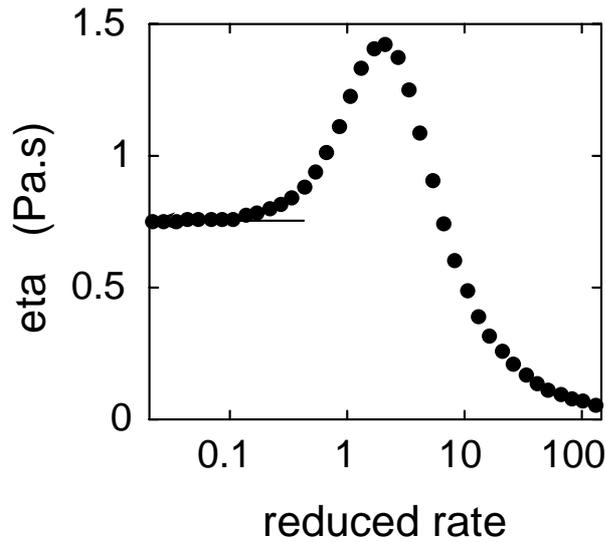

Figure 5): Viscosity versus reduced rate in steady shear for the sample with $\Phi = 13.7\%$ and $r = 3.4$. For this sample, $G(0) = 8\,Pa$ and $\tau_R = 0.135\,s$

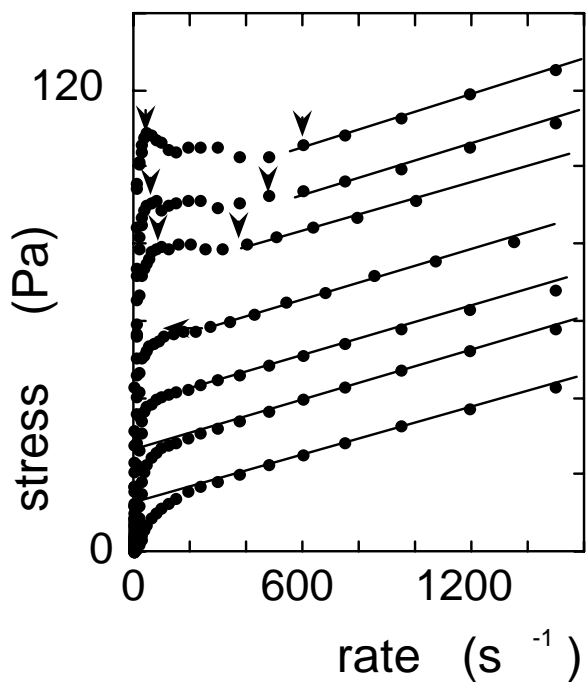

Figure 6): Stress versus rate in steady shear for samples with $\Phi = 13.7\%$ at increasing $r$'s. From the lower to the upper curve $r = 3.1, 3.2, 3.3, 3.4, 3.5, 3.6$ and $3.7$. The continuous straight lines are guides for the eyes. The arrows for the three upper curves show the limits of the unstable regime (see text).

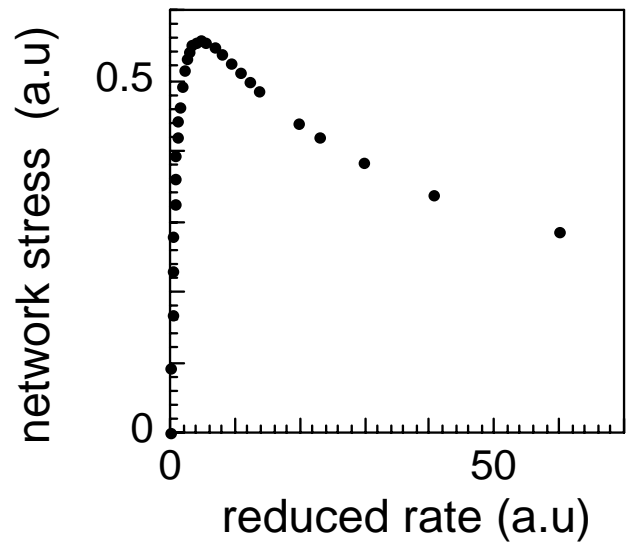

Figure 7): Network elastic stress versus the rate in arbitrary units as from expressions (12) and (13)

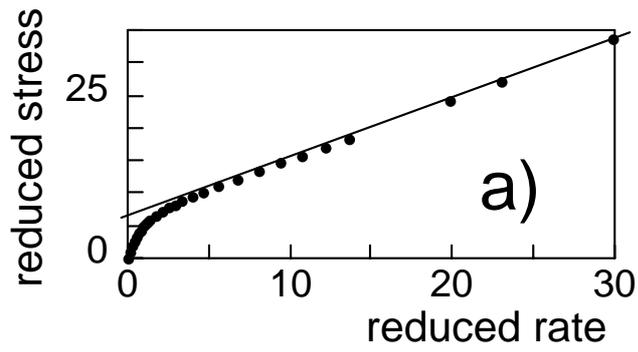

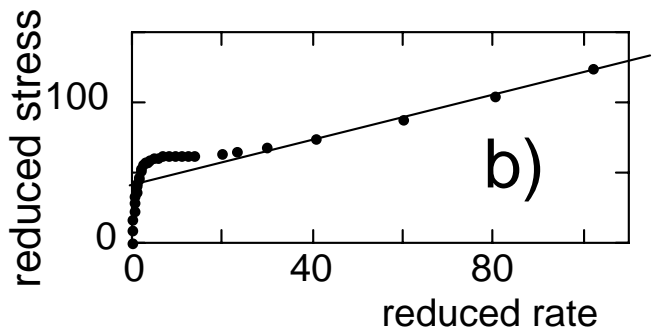

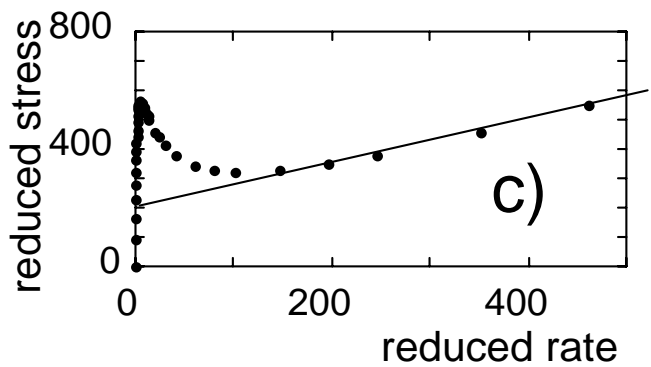

Figure 8): reduced total stress (elastic + viscous) versus the reduced rates all in arbitrary units. a), b) and c) correspond to increasing elastic to viscous relative contribution as state in the text.